\documentclass[apj,numberedappendix]{emulateapj}
\usepackage{graphics,epsf}
\usepackage{amsmath}                
\usepackage{amsfonts}               
\usepackage{amssymb}                
\usepackage{epsfig}
 \usepackage{epstopdf}
\usepackage{multirow}
\usepackage{graphicx}
\usepackage{float}
\usepackage{color}
\usepackage[para,online,flushleft]{threeparttable}

\def \kev{\rm{keV}}

\def \s{~\rm{s}} 
\def \km{~\rm{km}}

\def \erg{~\rm{erg}}

\def \yr{~\rm{yr}}

\def \keV{~\rm{keV}}

\newcommand{\nar}{{~\rm New Astronomy Reviews}}
\newcommand{\na}{{~\rm New Astronomy}}
\newcommand{\pasa}{{~\rm Publications of the Astronomical Society of Australia}}

\begin{document}

\title{Explaining the morphology of supernova remnant (SNR) 1987A with the jittering jets explosion mechanism}

\author{Ealeal Bear\altaffilmark{1} and Noam Soker\altaffilmark{1,2}}

\altaffiltext{1}{Department of Physics, Technion -- Israel Institute of Technology, Haifa
32000, Israel; ealealbh@gmail.com; soker@physics.technion.ac.il}
\altaffiltext{2}{Guangdong Technion Israel Institute of Technology, Shantou, Guangdong Province, China}
\begin{abstract}
We find that the remnant of supernova (SN)~1987A shares some morphological features with four supernova remnants (SNRs) that have signatures of shaping by jets, and from that we strengthen the claim that jets played a crucial role in the explosion of SN~1987A. Some of the morphological features appear also in planetary nebulae (PNe) where jets are observed. The clumpy ejecta bring us to support the claim that the jittering jets explosion mechanism can account for the structure of the remnant of SN~1987A, i.e., SNR~1987A.
We conduct a preliminary attempt to quantify the fluctuations in the angular momentum of the mass that is accreted on to the newly born neutron star via an accretion disk or belt. The accretion disk/belt launches the jets that explode core collapse supernovae (CCSNe).
The relaxation time of the accretion disk/belt is comparable to the duration of a typical jet-launching episode in the jittering jets explosion mechanism, and hence the disk/belt has no time to relax. We suggest that this might explain two unequal opposite jets that later lead to unequal sides of the elongated structures in some SNRs of CCSNe.
We reiterate our earlier call for a paradigm shift from neutrino-driven explosion to a jet-driven explosion of CCSNe.  
\end{abstract}


\section{INTROCUTION}
\label{sec:intro}
Supernova (SN) 1987A holds several puzzles. The first one was when the community had recognized it as the explosion of a blue giant. It seems that the merger of a binary companion with the SN progenitor can account for the explosion of a blue star (e.g., \citealt{Podsiadlowskietal1990, MenonHeger2017, Urushibataetal2018}) and for the asymmetrical explosion \citep{ChevalierSoker1989}. Furthermore, the strong binary interaction is likely to account for the presence of three rings (e.g., \citealt{MorrisPodsiadlowski2009}) that are observed around SN~1987A \citep{Wampleretal1990,Burrowsetal1995}.
The second puzzle is that observations have failed to detect any central compact object (CCO) remnant (e.g., \citealt{Haberletal2006, Indebetouwetal2014, McCrayFransson2016}). 
A third puzzle is the non spherical morphology of the explosion (e.g., \citealt{Kjaeretal2010, Abellanetal2017, Matsuuraetal2017}). 
  
Assuming that the delayed neutrino mechanism powered the explosion (e.g., \citealt{Jankaetal2016, Burrowsetal2017, OConnoretal2017, Muller2016} for recent papers on this mechanism, and \citealt{Papishetal2015} for problems with this explosion mechanism), some researchers (e.g., \citealt{Jankaetal2017, Wongwathanarat2017}) argue that instabilities can lead to such deviations from spherical symmetry in SN~1987A and similar supernovae, such as Cassiopeia~A. However, instabilities alone are not sufficient and a large scale asymmetrical outflow must take place, as \cite{Orlandoetal2016} argued for Cassiopeia~A.
By performing a detailed examination of the numerical simulations and the observations, \cite{Soker2017a} further concluded that the delayed neutrino mechanism cannot account for the asymmetrical elements distributions in SN~1987A and in Cassiopeia~A, and that jets can better explain the observations.  
 
Although the idea that jets can explode core collapse supernovae (CCSNe) is old (e.g., \citealt{Khokhlovetal1999}; for a recent review with references to earlier papers see \citealt{Soker2017a}), the understanding that jets might \textit{explode all CCSNe and that the jets operate in a negative jet feedback mechanism} is relatively new (e.g., \citealt{PapishSoker2011}; review by \citealt{Soker2016Rev}).
Our recent studies of supernova remnants (SNR; \citealt{BearSoker2017b, Bearetal2017, GrichenerSoker2017}) suggest that jets leave signatures in many of them, such as two opposite `ears', and that these signatures support the jet feedback mechanism. As SN~1987A is transforming now from the SN phase into the SN remnant phase, the study of its morphology in relation to SNRs can shed light on the explosion mechanism itself.
  
\cite{Wangetal2002} already suggested a jet-induced explosion model to account for the asymmetries in SN~1987A. According to their suggestion, the oxygen and calcium are expected to be concentrated in an expanding torus that shares the plane and northern blue shift of the inner circumstellar ring. However, the new observations of the SN~1987A ejecta \citep{Abellanetal2017} as we describe in section \ref{sec:1987A}, show that the torus-like ejecta and the inner circumstellar ring of SN~1987A do not share the same plane. For that, we set the goal of examining the morphology of SN~1987A in the frame of the jittering jets explosion mechanism. 

In the jittering jets explosion mechanism the pre-explosion core does not need to rotate fast, nor even mildly fast as required for example in the model of \cite{Wheeleretal2002}. 
Instead, stochastic variations of the specific angular momentum of the accreted gas on to the newly born neutron star (NS) lead to the formation of an intermittent accretion disk that launches jittering jets, e.g., intermittent jets with varying axis directions. 
The sources of the stochastic angular momentum are convective cells in the pre-explosion core \citep{GilkisSoker2014, GilkisSoker2016} and instabilities of the inflowing gas onto the NS \citep{Papishetal2015}, such as the spiral modes of the standing accretion shock instability (see, e.g., \citealt{BlondinMezzacappa2007, Rantsiouetal2011, Fernandez2015, Kazeronietal2017} for some papers on that instability).  
   
We conduct our study under the assumption that the jet feedback explosion mechanism, whether the jets jitter or not, accounts for all CCSNe, from energies of about few$\times 10^{50} \erg$ \citep{PapishSoker2011} and up to the most energetic CCSNe \citep{Gilkisetal2016}, even in cases where a magnetar is formed (e.g., \citealt{Soker2016Mag1, Chenetal2017, SokerGilkis2017}).

We structure the paper as follows. In section \ref{sec:Clump} we discuss the morphology of four SNRs, and in section \ref{sec:1987A} we focus on SNR~1987A. 
In section \ref{sec:jittering} we list the key properties of these SNRs, and then explain them in the frame of the jittering jets explosion mechanism. 
Readers who are interested mainly in the jittering jets explosion mechanism can skip sections \ref{sec:Clump} and \ref{sec:1987A} and move directly to \ref{sec:jittering}. 
In section \ref{sec:PNe} we use some hints from planetary nebulae (PNe) on shaping by jets. 
We summarize in section \ref{sec:summary}. 

\section{Clumps and the jet direction}
\label{sec:Clump}
 
We start by reviewing the structure of several SNRs, concentrating on clumps and jets. 

\subsection{Cassiopeia~A}
\label{subsec:Cas A}
 
\begin{figure*}
 \centering
 \hskip -1.00 cm
 \includegraphics[trim= 0.0cm 2.0cm 0.0cm 1.5cm,clip=true,width=0.85\textwidth]{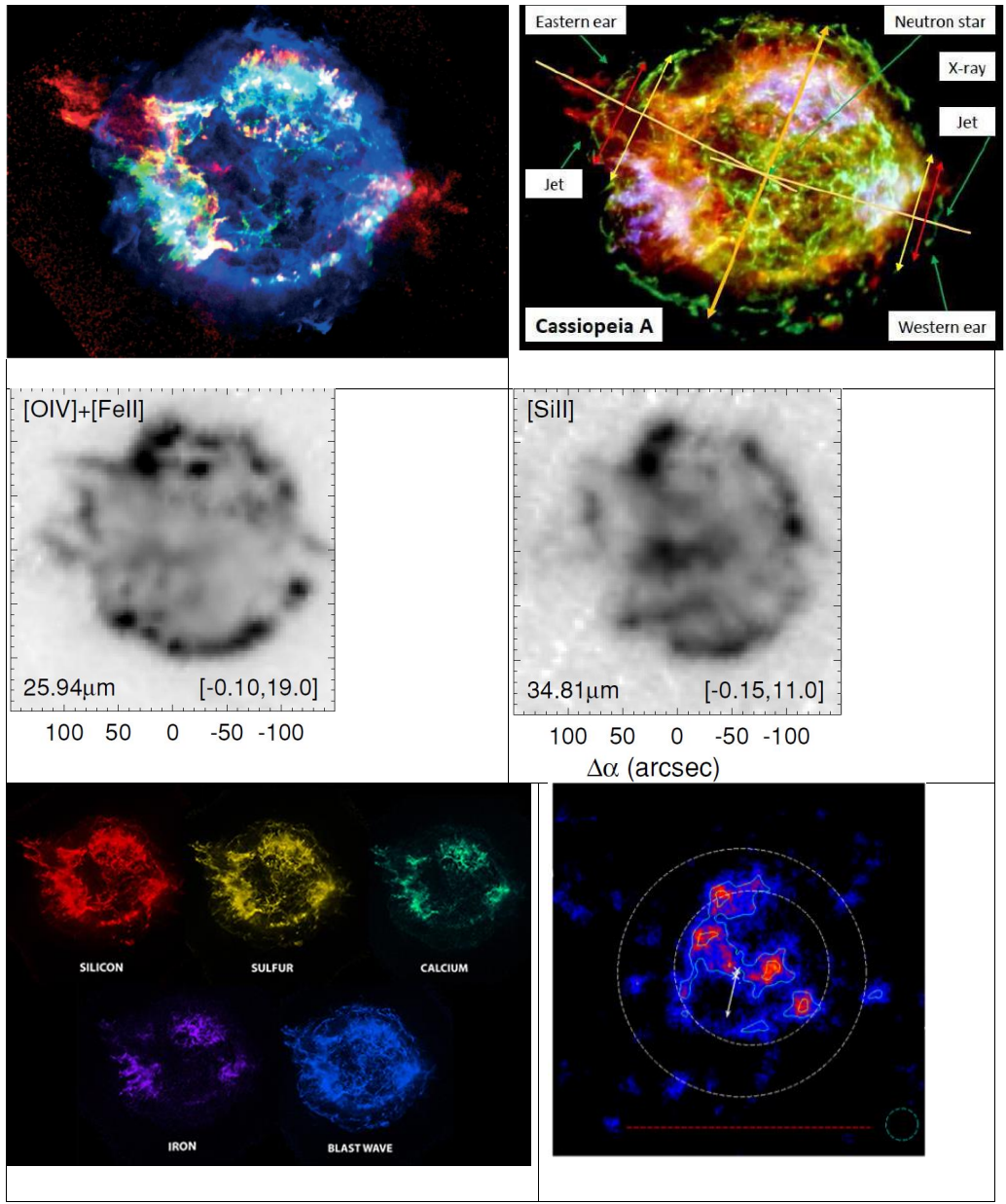}\\
 \vskip -1.00 cm
 \caption{Images of SNR Cassiopeia~A that emphasize the clumpy nature of the remnant. The upper-left panel is a three-color image showing the jets location (taken from \citealt{Schureetal2008}). Green represents X-ray emission of Si~{\sc xiii} and the jet image is obtained by the ratio of Si~{\sc xiii} to Mg~{\sc xi} X-ray line emission. The upper-right panel is from \cite{GrichenerSoker2017} who added notation on an X-ray image taken from the Chandra gallery (based on \citealt{Hwangetal2004}). Red, blue and green represent Si He$\alpha$ ($1.78-2.0 \keV$), Fe K ($6.52-6.95 \keV$), and $4.2-6.4 \keV$ continuum, respectively. The two panels in the middle row are taken from \cite{Smithetal2009} and show surface brightness maps in the emission lines indicated (for other emission line - surface maps, see \citealt{Smithetal2009}); the y-axis is in the same scale of arcsec as the x-axis in both panels. The lower-left panel presents the distribution of some elements (Credit: NASA/CXC/SAO). The lower-right panel shows the Ti distribution in Cassiopeia~A, with circles that mark the forward and reverse shocks (from \citealt{Jankaetal2017} based on \citealt{Grefenstetteetal2014}).  }
 \label{Clumps_CasA}
 \end{figure*}

We present the relevant observations of Cassiopeia~A in Fig. \ref{Clumps_CasA}. 
 \cite{GrichenerSoker2017} identified two opposite protrusions that are termed `ears' in the image of Cassiopeia~A (upper-right panel where they added notation on an image based on \citealt{Hwangetal2004}), and marked some properties of the ears, e.g., the diameter of the base of the ear on the SNR main shell (yellow double-headed arrow) and the direction from the center (yellow lines) that they identify as the direction of the jets that inflated the ears. One very relevant property of Cassiopeia~A to the present study is that the two ears are not equal, as the eastern jet is much larger than the western one. 

The upper left panel that is taken from \cite{Schureetal2008} clearly shows the eastern jet and the clumpy nature of he remnant. The two middle-row panels, that we take from figure 2 of \cite{Smithetal2009}, further emphasize the clumpy nature of the remnant. 
The lower left panel that we take from Chandra website presents the different metal distributions. We particularly point to the presence of two large iron clumps away from the jets' axis and  to the fact that some elements, such as Si, have clumps that coincide with the jet direction and other elements, such as Fe, have a different distribution. 
We show the titanium distribution of Cassiopeia~A that we take from \cite{Grefenstetteetal2014} in the lower-right panel of Fig. \ref{Clumps_CasA}. The titanium distribution close to the center seems to be aligned with the jets' axis, while further out there are clumps mis-aligned with the jets' axis. 

\cite{Schureetal2008} and \cite{Lamingetal2006} discuss jet-driven explosion mechanism for Cassiopeia~A. \cite{Schureetal2008} point out that even if the SNR is spherically shaped the explosion can be accompanied by jets.  
In the context of the jittering jets explosion mechanism the ears of Cassiopeia~A were shaped by the last jets-launching episode \citep{GrichenerSoker2017}.

The main points to take from Cassiopeia~A to our study of SNR~1987A are that (1) the two opposite ears are highly non-equal, and (2) that clumps have no unique sense of asymmetry.  
\subsection{Vela}
\label{subsec:Vela}

\begin{figure}
 \centering
 \hskip 2.00 cm
 \includegraphics[trim= 0.0cm 0.0cm 0.0cm 0.0cm,clip=true,width=0.85\textwidth]{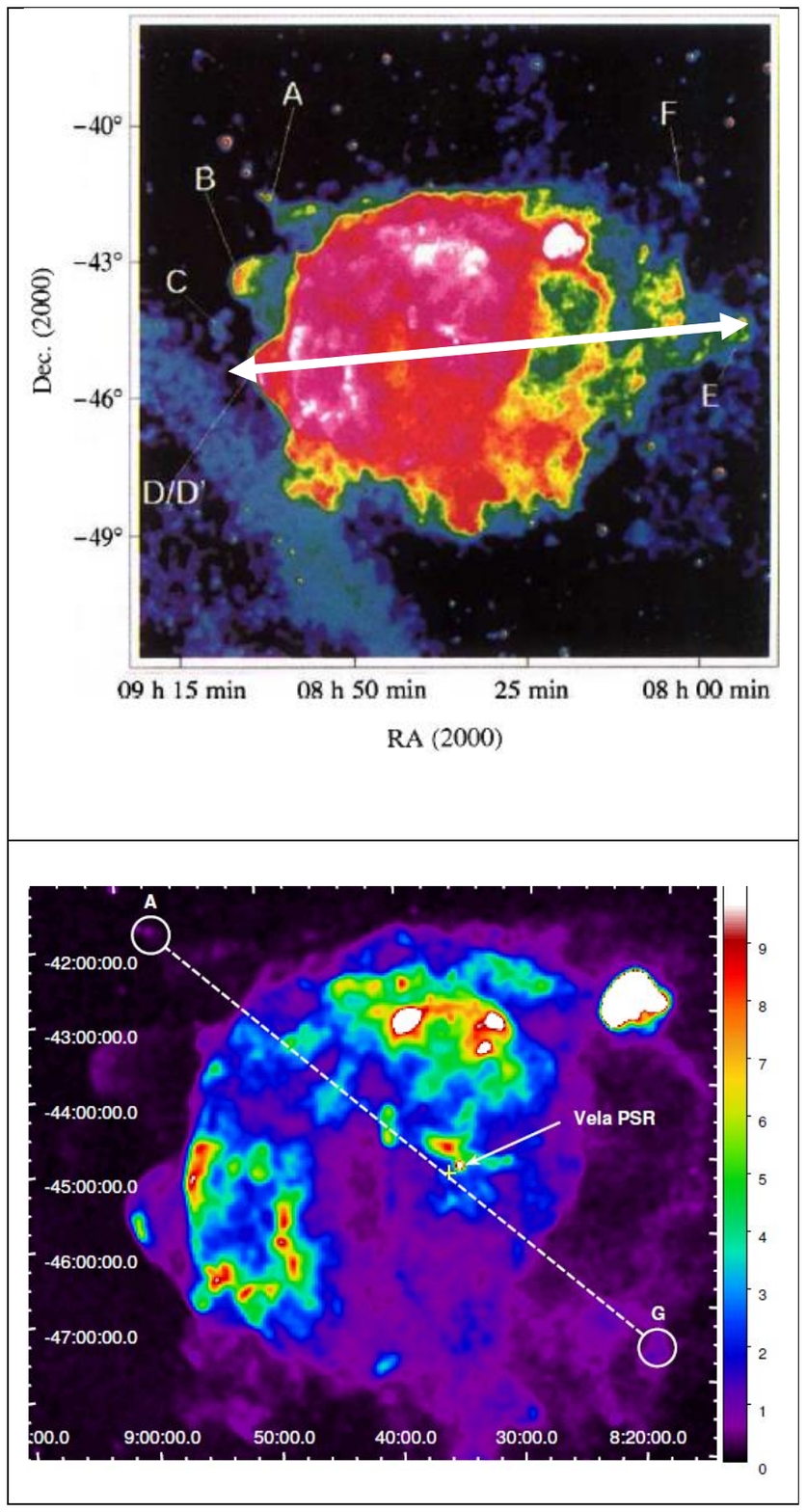}\\
 \vskip -6.50 cm
 \caption{Images of the SNR Vela that emphasize clumps and possibly jets. 
The {upper panel} is taken from \cite{Aschenbachetal1995}. It is a $0.1 - 2.4 \keV$ ROSAT all-sky survey image where the colors represent the surface brightness, from light blue, yellow, red and white at an increasing intensity.
We added our proposed jets' axis by a double-head white arrow according to \cite{GrichenerSoker2017}.
The {lower panel} is taken from  \cite{Garciaetal2017}. It is a $0.44-2.01 \kev$ ROSAT All-Sky Survey image. They inferred and marked the explosion site by a yellow cross. They argue for two opposite jets along the line connecting shrapnel A and G as they marked with a dotted line.
}
 \label{Clumps_Vela}
 \end{figure}

Another clumpy SNR that is relevant to our study is Vela, that we present in Fig. \ref{Clumps_Vela}.    
In the upper panel we present an image from \cite{GrichenerSoker2017}, who marked with a double-head arrow their proposed two opposite jets (the image is based on the $0.1 - 2.4 \keV$ ROSAT all-sky survey taken from \citealt{Aschenbachetal1995}). \cite{GrichenerSoker2017} assume that each of the two opposite ears was inflated by a jet. 
\cite{Garciaetal2017}, on the other hand, suggest that two opposite jets formed the Si clumps that they marked by A and G (lower panel of Fig. \ref{Clumps_Vela}). Based on the ears in Cassiopeia~A, we adopt the view that the jets' axis is as taken by \cite{GrichenerSoker2017}. We do point out that according to the jittering jets explosion mechanism in some cases the last two jet-launching episodes can leave imprints on the SNR (and not only the last one), and they might be along different directions. 

\subsection{G$292.0+1.8$}
\label{subsec:G292018}
   
\begin{figure*}
 \centering
 \hskip -0.00 cm
 \includegraphics[trim= 0.0cm 0.0cm 0.0cm 0.0cm,clip=true,width=0.90\textwidth]{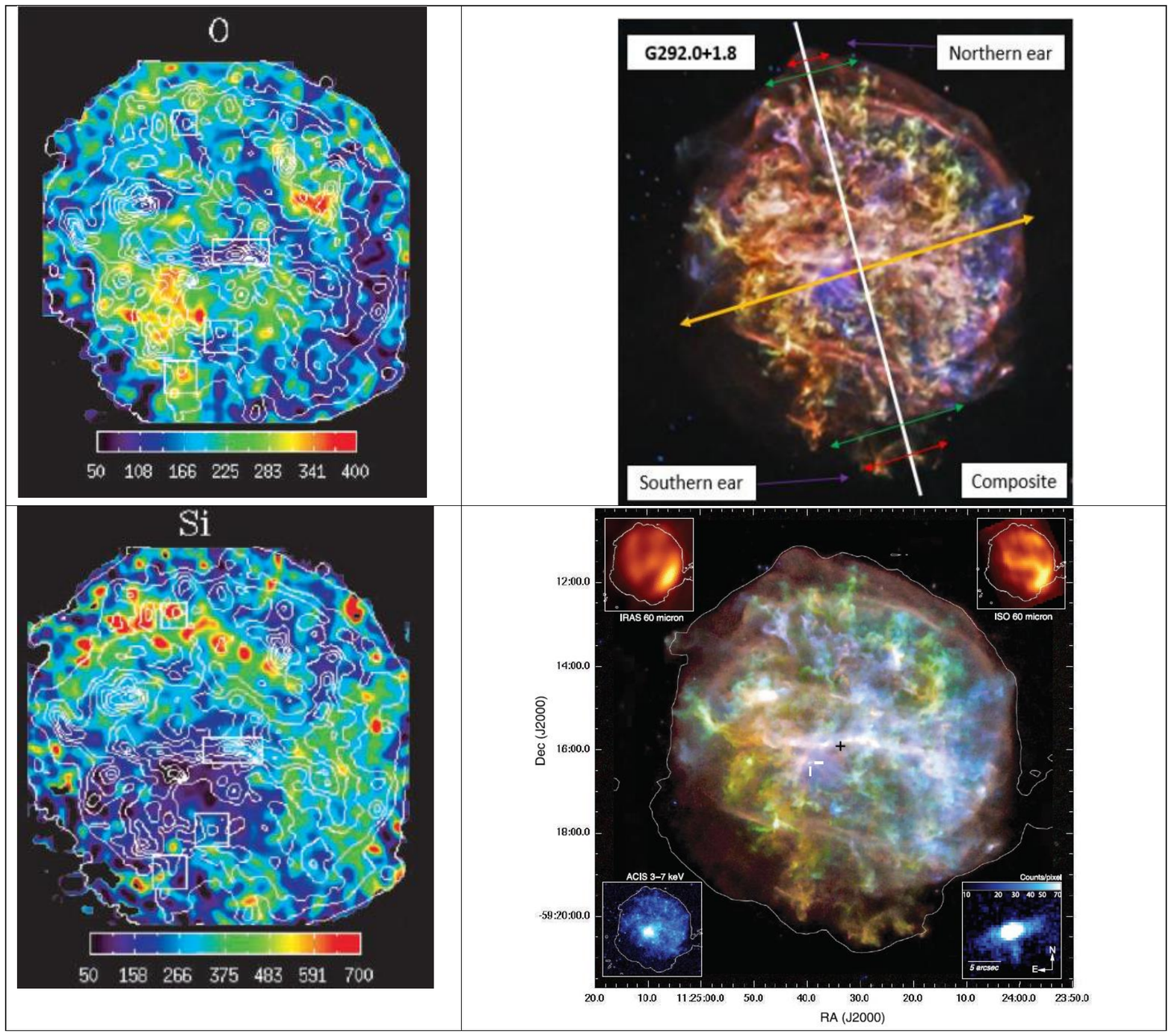}\\
 \vskip -7.00 cm
 \caption{The clumpy ejecta of SNR G$292.0+1.8$ and the proposed jets. The {left panels} present the clumpy morphology of oxygen and silicon as indicated (from \citealt{Parketal2002}). It is the equivalent width images for the elemental species O and Si.
The {upper-right panel} is from our earlier paper \citep{Bearetal2017}. The almost vertical white line shows our proposed jets' axis. It is a composite image taken from the Chandra gallery, where red, orange, green and blue colors represent O~Ly$\alpha$ combined with Ne~He$\alpha$ ($0.58 - 0.71$ and $0.88 - 0.95 \keV$), Ne~Ly$\alpha$ ($0.98 - 1.10 \keV$), Mg~He$\alpha$ ($1.28-1.43 \keV$), and Si~He$\alpha$ combined with S~He$\alpha$ ($1.81 - 2.05$ and $2.40 - 2.62 \keV$), respectively. White represents the optical band. 
The {upper-right panel} is based on the lower-right panel (taken from \citealt{Parketal2007}).
In the {lower-right panel} red represents O~Ly$\alpha$ and Ne~He$\alpha$, orange represents Ne~Ly$\alpha$, green represents Mg~He$\alpha$, and blue represents Si~He$\alpha$ and S~He$\alpha$.}
 \label{Clumps_G292018}
 \end{figure*}

Another SNR with clumpy ejecta and signatures of jets is G$292.0+1.8$, that we present in Fig. \ref{Clumps_G292018}. 
The two left panels that we take from \cite{Parketal2002} emphasize the oxygen (upper-left) and silicon (lower-left) clumpy morphology of the ejecta.
The jets' axis as we proposed in an earlier paper \citep{Bearetal2017} based on the location of the two opposite ears is marked by an almost vertical white line on the upper right panel. As in other cases, we assumed there that each ear is inflated by one jet. The image itself is based on an image by \cite{Parketal2007} that we show in the lower right panel. 
The green and red double-arrows in the upper right panel, indicate properties of the ears \citep{Bearetal2017}.

The relevant properties to take from SNR G$292.0+1.8$ is that the clumps in different metals have different morphologies, and the metals not necessarily are concentrated along the jets' axis. The upper-left panel of Fig. \ref{Clumps_G292018} shows that the two bright oxygen peaks are on the two sides of the jets' axis. The clumpy silicon distribution (lower-left panel) has a different morphology, but it also seems to extend more or less perpendicular to the jets' axis.

\subsection{W49B}
\label{subsec:W49B}
  
\begin{figure}
 \centering
\hskip -2.00 cm
 \includegraphics[trim= 2.4cm 0.0cm 0.0cm 0.0cm,clip=true,width=0.65 \textwidth]{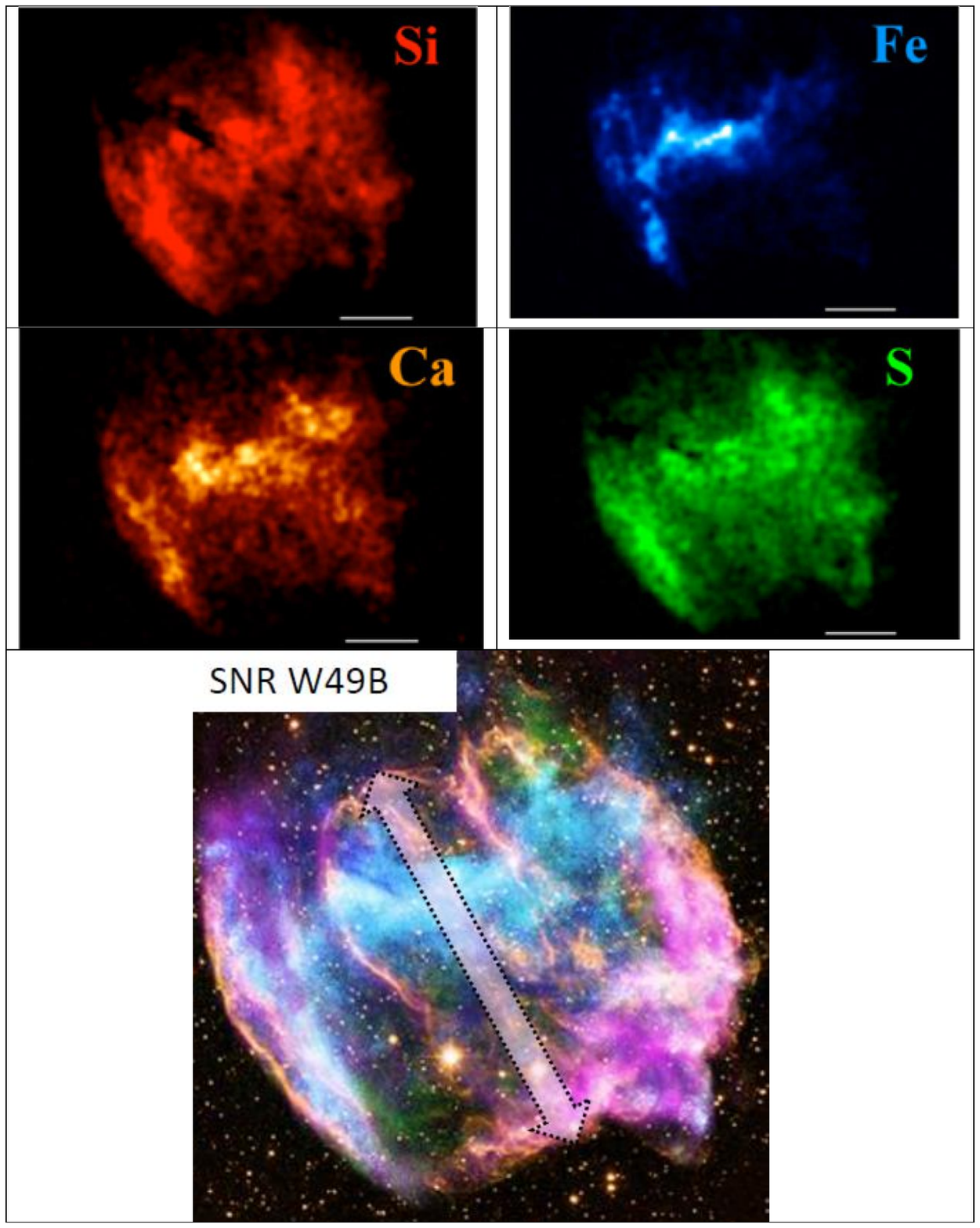} \\
 \vskip -3.50 cm
 \caption{The barrel-shaped SNR W49B. The {four upper panels} of SNR W49B are taken from \cite{Lopezetal2013}, and are the continuum-subtracted images of Si~{\sc xiii} and Si~{\sc xiv} (Si), of S~{\sc xv} and S~{\sc xvi} (S), of Ca~{\sc xix} and Ca~{\sc xx} (Ca), and of Fe~{\sc xxv}. 
The {lower panel} is taken from \cite{BearSoker2017b} where the jets' axis is marked by a thick double-head arrow. X-ray is presented in green and Blue, Infrared in yellow, and Radio in magenta.}
 \label{W49B}
 \end{figure}

The SNR W49B is different than the previous three SNRs in that it has no ears. In the four left panels of Fig. \ref{W49B} we present X-ray images taken with the Chandra X-ray Observatory by \cite{Lopezetal2013}. In an earlier paper \citep{BearSoker2017b} we compared the morphology of W49B with the morphologies of several PNe, and pointed out similar features such as the barrel shape and the 'H' shape, that in PNe are taken to be formed by jets. We then used this comparison to conclude that jets shaped SNR W49B, and that the jets propagated along the symmetry axis of the barrel, as we indicate in the lower panel of Fig. \ref{W49B} by the thick double-head arrow.
It is interesting to note that \cite{ZhouVink2017} suggest that SNR W49B is a type Ia SNR. We here consider it to be a SNR of a CCSN. 
 
Two types of scenarios were proposed to explain the barrel shaped structure of W49B. The first is a result of an inhomogeneous ISM interactions and the second is a jet-driven explosion of a massive star \citep{Lopezetal2013}. The direction of the jets however, is still under debate. We suggest \citep{BearSoker2017b} that the jets' axis is along the symmetry axis of the barrel, while \cite{Lopezetal2013} take the jets' axis along the dense iron bar. 
These two directions are perpendicular to each other. In our proposed scenario the ears that the jets inflated in SNR W49B were already dispersed, and hence the gas that was in the ears is too faint to be observed.
  
The properties to take from W49B is that a concentration of iron, calcium and some other elements can be in a bar that is perpendicular to the jets' axis. 
  
Beside the above four SNRs, prominent clumpy morphologies are observed in the Crab Nebula (e.g., \citealt{Satterfieldetal2012}), in Puppis A (e.g., \citealt{Katsudaetal2010}), in G$180.0−1.7$ (S147; e.g., \citealt{Jeongetal2012}), in G109.1-1.0 (CTB 109; e.g., \citealt{SnchezCrucesetal2018, Sasakietal2013}) and in Kes~75 (e.g., \citealt{Vink2012}).

\section{SNR~1987A}
\label{sec:1987A}

\begin{figure}
 \centering
 \hskip -14.00 cm
 \vskip -1.90 cm
\includegraphics[trim=2.3cm 3.0cm 0.0cm 0.0cm,clip=true,width=0.96\textwidth]{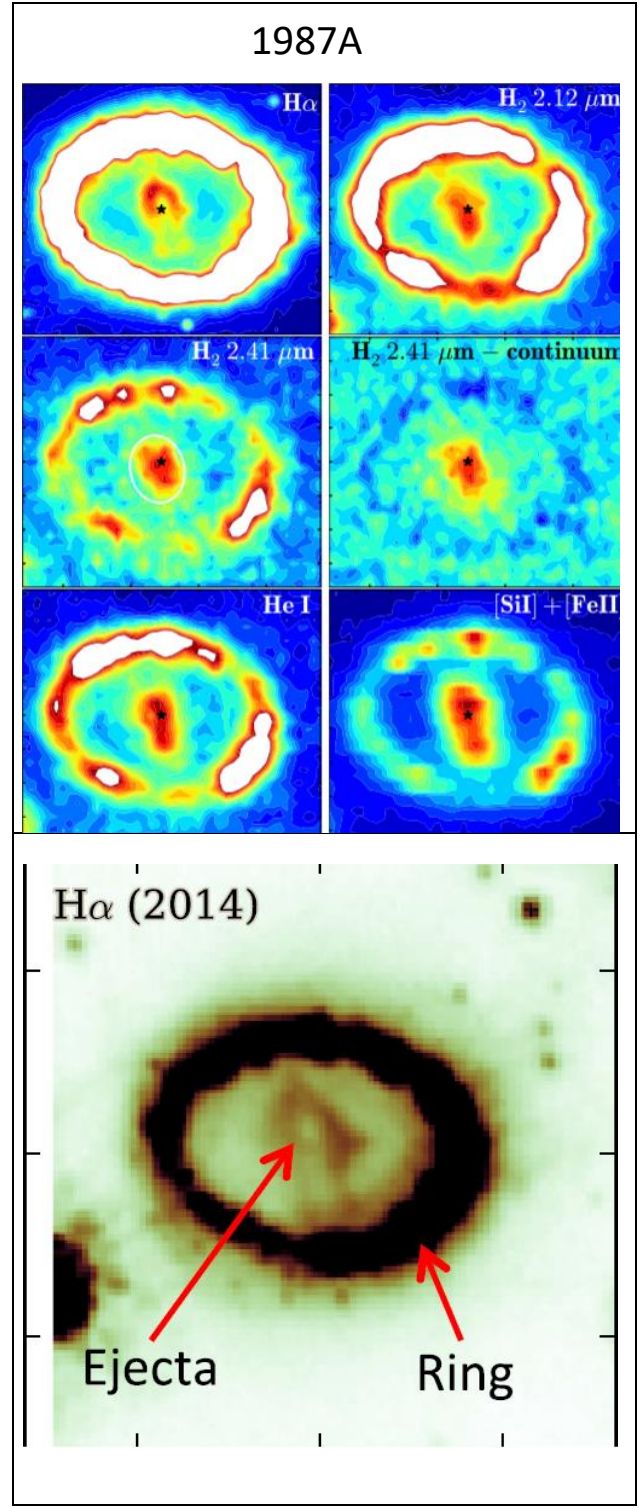}\\  
 \vskip -0.00 cm
 \caption{The relative structures of the ejecta and the equatorial ring of SNR~1987A. The {upper six panels} are taken from \cite{Franssonetal2016}. They show the spatial intensity distribution in different lines and bands as indicated in each panel. Blue represents minimum and red represents maximum intensity.  
The {lower panel} is taken from \cite{Matsuuraetal2017} and shows H$\alpha$ image from the Hubble Space Telescope.}
 \label{fig:1987A_Xray}
 \end{figure}

\begin{figure}
 \centering
 \hskip -0.700 cm
 \includegraphics[trim= 0.0cm 0.0cm 3.9cm 0.0cm,clip=true,width=0.51\textwidth]{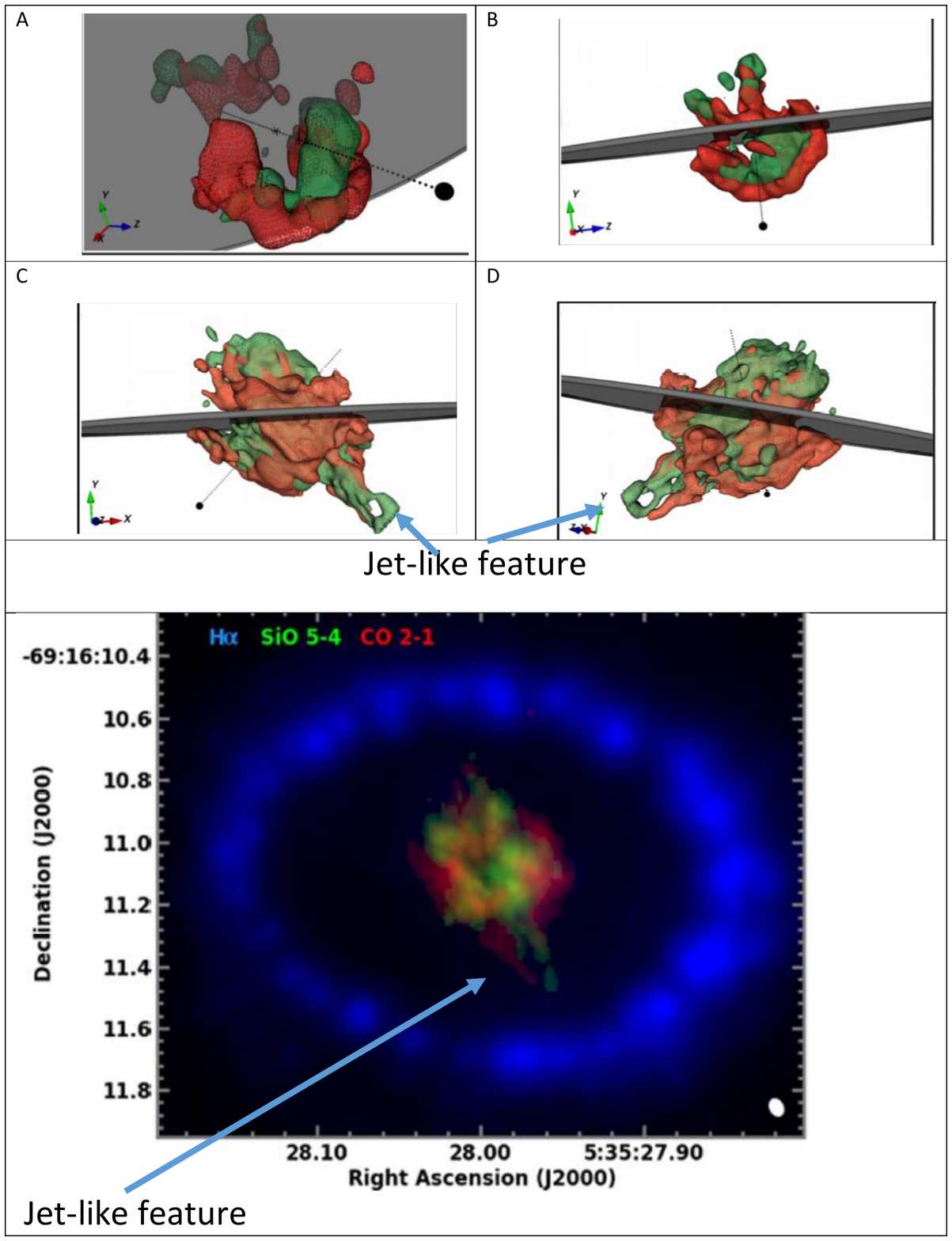}\\
 \vskip -2.00 cm
 \caption{The structure of the molecular gas in SNR~1987A, taken from \cite{Abellanetal2017}. We added arrows pointing at what we identify as a morphological feature that formed by a jet. 
 The {upper panels} show 3D view of cold molecular emission, CO in red and SiO in green. The line of sight and the position of the observer are marked by the black dotted line and black filled sphere respectively. The gray plane is the equatorial plane. Panel A presents emission at $60\%$ of peak emission, panels B and C are for $30\%$ of peak emission, and panel D is for $50\%$ of peak emission.
The {lower panel} is taken from \cite{Abellanetal2017} and shows the molecular emission and H$\alpha$ emission from SNR 1987A. The compact part in the center corresponds to the peak intensity observed with ALMA, where red is CO $2-1$ and  green is SiO $5-4$. The blue ring corresponds to H$\alpha$ emission.}
 \label{fig:1987A}
 \end{figure}

We start by describing the relevant morphological properties of the ejecta of SNR~1987A and its relation to the equatorial ring that was blown by the star $\approx 2 \times 10^4 \yr$ before the explosion (e.g., \citealt{Meaburnetal1995}).    
In Fig. \ref{fig:1987A_Xray} we present some images out of many that exist in the literature (e.g., \citealt{Franssonetal2015, Franssonetal2016, Larssonetal2016}). 
In Fig. \ref{fig:1987A} we present results from \cite{Abellanetal2017} who study and discuss the structure of the molecular gas (SiO and CO) as well as of metals ([Si {\sc i}] and [Fe {\sc ii}]) in the ejecta. 

\cite{Abellanetal2017} identify a CO emission torus-like shape perpendicular to the equatorial ring, and argue that this is an evidence of asymmetrical explosion. The SiO emission presents a clumpier structure and is concentrated in a broken shell rather than in a torus. There is a bright SiO emission blob below the equatorial plane (defined by the ring).
\cite{Larssonetal2016} find that the infrared [Si {\sc i}]+[Fe {\sc ii}] emission is concentrated in two asymmetric lobes close to the plane of the ring.  
According to \cite{Abellanetal2017} the structure of the molecular emission is brighter closer to the center than the [Si {\sc i}]+[Fe {\sc ii}] emission is, and it is not aligned with the [Si {\sc i}]+[Fe {\sc ii}] structure (the two lobes).

\cite{Matsuuraetal2017} studied the distribution of molecules in SNR~1987A and argue that Rayleigh-Taylor instabilities caused the mixing of elements in the ejecta. \cite{Abellanetal2017} also discuss the clumpy ejecta, and compare to expectations of neutrino driven explosions. They find that none of these models fit all observations. Several months earlier \cite{Soker2017a} already concluded that the instabilities in neutrino driven explosions cannot account for the structure of the SNR~1987A ejecta, in particular to the two emission [Si {\sc i}]+[Fe {\sc ii}] lobes. We here adopt the result of \cite{Soker2017a}, that was later confirmed by \cite{Abellanetal2017}, that a neutrino driven explosion cannot explain the morphology of the ejecta, as well as his conclusion that to account for the morphology of the ejecta of SNR~1987A jets must be introduced in addition to the instabilities.  
 
There are earlier suggestions that SNR~1987 was driven by jets, e.g., \cite{Wangetal2002}. They note that the ejecta symmetry axis is inclined by about $15^\circ$ to the axis of the CSM rings, but still assume that the CSM rings and the ejecta share the same symmetry axis. This cannot be the case any more, as the symmetry axis of the molecular torus, the [Si {\sc i}]+[Fe {\sc ii}] blobs, and the molecular elongate structure deviate substantially from the symmetry of the CSM rings. 
\cite{Nagataki2000}, calculates the nucleosynthesis that is expected in a jet driven explosion. Based on these calculations he suggests that the jets did not share the same symmetry axis as the CSM rings, and that the two jets were not equal in their intensity.
We argue below that the jittering jets mechanism can account for these asymmetries. 

\cite{Kjaeretal2010} attribute the non-symmetric explosion that they deduce from their observations to instabilities alone. They argue that the elongation of the ejecta in the plane of the equatorial ring argues against a jet-induced explosion due to stellar rotation. The jittering jets explosion mechanism accounts for misalignment of the symmetry axis of the equatorial ring and of the direction of elongation, as in the jittering jets explosion mechanism the momentarily direction of the two opposite jets can be highly inclined to the average angular momentum axis. We therefore agree with \cite{Kjaeretal2010} that the directions of the jets in the case of SNR~1987A cannot be determined only by the stellar rotation, but disagree with them that instabilities alone can explain the morphology of SNR~1987A. 
 
\section{The jittering jets explosion mechanism}
\label{sec:jittering}

\subsection{Key properties}
\label{subsec:Properties}

We discuss here only the prominent properties of SNR~1987A that are relevant to us, and emphasize the relation to the four SNRs that we describe in section \ref{sec:Clump}. In doing so we actually refer to SN~1987A as a SNR.  
 
(1) The ejecta of SNR~1987A is clumpy and possesses a large departure from spherical symmetry, in particular an elongated morphology. The closest SNR to this structure out of the four that we study in section \ref{sec:Clump} is Vela (Fig. \ref{Clumps_Vela}). As shown by \cite{Soker2017a} instabilities alone in the frame of the neutrino-driven explosion cannot account for this structure. Jets are required. 

(2) The elongation of the ejecta is not perpendicular to the equatorial ring. In the other SNRs we simply have no observations of the circumstellar matter (CSM) before explosion, and hence have no indication of the pre-explosion CSM symmetry.  This implies that jets can be launched along an axis that is not directly correlated with the angular momentum axis of the stellar progenitor of the CCSN. In the context of the jittering jets explosion mechanism this implies that the angular momentum fluctuations of the material that is last to be accreted on to the newly born NS are very large. 

(3) The two opposite elongated parts seen in molecules are not equal. The departure from mirror symmetry of Cassiopeia~A with its unequal ears (jets) is the closest to the unequal elongated sides in SNR~1987A. In the frame of the jet-driven explosion mechanism this implies unequal opposite jets. 
  
We turn now to discuss these key properties in the frame of the jittering jets explosion mechanism.

\subsection{Large angular momentum fluctuations}
\label{subsec:fluctuations}
    
There are two indications to large asymmetrical explosions of CCSNe, the metal distribution and the NS natal kick. 
The highly non-spherical distributions of metals in the five SNRs that we discussed in previous sections show that the explosion close to the center, where Fe, Si, O, etc. are formed, is highly non-spherical. 
  
In the accelerating mechanism of the NS that we adopt here, and is termed the gravitational tug-boat mechanism \citep{Nordhausetal2010, Janka2017}, one or more dense clumps that are expelled by the explosion, gravitationally attract the NS and accelerate it. 
The gravitational tug-boat mechanism is a relatively long-duration process lasting several seconds after accretion has ended, and when the dense regions are accelerated from about $100 \km$ to $\approx 10^4 \km$ from the origin \citep{Nordhausetal2010, Wongwathanaratetal2013, Janka2017}.
The point here is that the instabilities start to develop inside the stalled shock. The development of instabilities and clumps are accompanied by large vortices behind the stalled shock, i.e., at $\approx 100 \km$ from the newly born NS (e.g., \citealt{Nordhausetal2010}).
The departure of the net momentum of the ejecta from spherical symmetry is about $\alpha_{\rm ej} \simeq 0.1$, where 
\begin{eqnarray}
\begin{aligned}
\alpha_{\rm{ej}} = 
\left(  \biggl|  \int \vec{p}_{\rm ej} d m_{\rm ej} \biggr|  \right) 
\left( \int \vert \vec{p}_{\rm ej} \vert d m_{\rm ej} \right)^{-1},
\end{aligned}
\label{eq:alphaej}
\end{eqnarray}
$\vec{p}_{\rm ej}$ is the momentum per unit mass of ejected mass elements, and the integration is over the ejecta mass that influences the NS natal kick, $m_{\rm ej} \approx 0.1 M_\odot$.
 
Based on vortices that were found in the simulation of \cite{Nordhausetal2010} we assume that at $r \simeq 100 \km$ the accreted mass on to the newly born NS has similar typical fluctuations, not only in the radial direction but also in the transverse direction.  This implies a net specific angular momentum of $\alpha_{\phi} \equiv j_z/j_{\rm Kep} (100 \km) \simeq \alpha_{\rm ej} \simeq 0.1$, where $j_z$ is the typical amplitude of fluctuating specific angular momentum of the accreted mass and $j_{\rm Kep} (100 \km)$ is the specific angular momentum of a circular orbit at $r=100 \km$. We take $z$ to be the axis of the angular momentum.  
 If the accreted gas conserves its angular momentum, then above the surface of the NS, about $20 \km$, the ratio of fluctuating angular momentum to that of a Keplerian circular orbit will be 
\begin{eqnarray}
\begin{aligned}
\frac{j_z}{j_{\rm Kep-NS}} \simeq 2.2 \alpha_\phi \simeq 0.2.
\end{aligned}
\label{eq:alphatheta}
\end{eqnarray}

As the material possesses a specific angular momentum of $j_z \ne 0$, the direction from which the mass flows on to the NS is limited to an angle of $\theta > \theta_a$, where $\theta$ is measured from the $z$-axis.
This actually implies that the density of the inflow inside the two cones around the polar axis, $\theta < \theta_a$, is very low.   

This limiting angle for accretion, $\theta_a$, is given by the balance between the gravitational forces and the centrifugal force \citep{PapishGilkisSoker2015}
\begin{equation}
\theta_a = \sin^{-1} \sqrt{\frac{j_z}{j_\mathrm{Kep-NS}}} \simeq 30^\circ \left( \frac{\alpha_\phi}{0.1}\right)^{1/2}.
\label{eq:alphaA}
\end{equation}
Even that the flow is sub-Keplerian, and hence at the beginning forms an accretion belt rather than an accretion disk on the surface of the neutron star, earlier studies suggest that the belt can launch jets \citep{SchreierSoker2016}.   
 
Using the results of 3D simulations of the standing accretion shock instability (SASI) from \cite{Fernandez2010}, \cite{PapishGilkisSoker2015} found the limiting angle to have typical values up to $\theta_a ({\rm SASI}) \simeq 12 ^\circ$. We argue from the careful examination of the asymmetrical structures of five SNRs, and in particular that of SNR~1987A, that the specific angular momentum fluctuations are larger than what \cite{Fernandez2010} finds in his simulations of the SASI.
Probably three processes or more contribute together to the large fluctuations in the angular momentum of the accreted gas. These are initial perturbations in the convective zone of the pre-collapse core \citep{GilkisSoker2014}, the SASI (e.g., \citealt{Fernandez2010, Fernandez2015}), and turbulence driven by neutrino heating (e.g., \citealt{Kazeronietal2018}). 

In a recent study we \citep{BearSoker2018} found that the direction of the NS natal kick and our inferred jets' axis in SNR tend to avoid alignment. The signature of jets in most SNRs come only from the last (or last two) jets-launching episode in the jittering jets explosion mechanism \citep{Bearetal2017}.  
This misalignment supports the assumption we take here that the density fluctuations that lead to the NS natal kick are related to the fluctuations that feed the intermittent accretion disk around the newly born NS. This disk launches the jets along its (temporary) angular momentum axis. 

The larger specific angular momentum fluctuations we argue for here, make more likely the earlier assumption of the jittering jets explosion mechanism that this type of accretion flow can form an intermittent accretion disk/belt/torus around the newly born NS. Furthermore, this belt is flatter (larger limiting angle from the pole $\theta_a$), and hence more likely to launch jets. 

\subsection{Asymmetrical double jets}
\label{subsec:asymmetryjets}

Let us examine property (3) in section \ref{subsec:Properties}. The unequal ends of the elongated structure of SNR~1987A and the unequal sizes of the two opposite jets in Cassiopeia~A require that the two opposite jets that shaped these structures, most likely from the last jet-launching episode, are not equal. 
  
Consider a temporary disk that is forming around the newly born NS because of angular momentum fluctuations (section \ref{subsec:fluctuations}). The time it takes the disk to settle to (semi) steady state structure at radius $R_{\rm d}$ is about the viscosity time scale of the disk at that radius, that is given by (e.g., \citealt{Dubusetal2001})
\begin{eqnarray}
\begin{aligned}
t_{\rm{visc}} \simeq \frac{R_{\rm d}^2}{\nu}
\simeq 0.06 &
\left(\frac{R_{\rm d}}{20 \km} \right)^{3/2} 
\left(\frac{\alpha}{0.03}\right)^{-1}
\\ & \times
\left(\frac{R_{\rm d}}{3H} \right)
\left(\frac{v_{\phi}}{3C_s} \right)  \s
\end{aligned}
\label{eq:tvisc1}
\end{eqnarray}
where $\nu=\alpha C_s H$ is the viscosity of the disk parameterized with the disk$-\alpha$ parameter,
$H$ is the vertical thickness of the disk, $C_s$ is the sound speed, and $v_\phi$ is the Keplerian velocity. 
We scale the parameters for an initial not-so-thin disk, and for a NS mass of $1.4 M_\odot$.  
The orbital time at radius $R_{\rm d}$ is 
$2 \pi R_{\rm d}/v_\phi = 0.0013 (R_{\rm d}/20 \km)^{3/2} \s$.
 
 The viscous time scale given in equation (\ref{eq:tvisc1}) is about the life time of one jet-launching episode in the jittering jets explosion mechanism, $t_{\rm jet} \simeq 0.05-0.1 \s$ \citep{PapishSoker2011, PapishSoker2014a, PapishSoker2014b}. This implies that during one jet-launching episode the disk has no time to settle down to a complete steady state. One plausible outcome is that the two sides of the disk are not equal, and hence the two opposite jets are not equal.  
This is our explanation for the unequal two opposite jets. 

\section{Lessons from planetary nebulae}
\label{sec:PNe}

In section \ref{subsec:Properties} we claimed that the ejecta of SN~1987A was shaped by jets that changed their symmetry axis during the explosion. A comparison to PNe might shed light on our claim. 
In the past we \citep{Bearetal2017, BearSoker2017b} compared the morphologies of some SNRs that possess axi-symmetrical features (e.g., W49B; G$292.0+1.8$; RCW~103) to PNe. Such morphological features include two opposite small protrusions (termed ears), equatorial mass concentration, and a faint region along a symmetry axis.   
We now consider cases where there is no symmetry at all, namely, no axi-symemtri, no mirror-symmetry, and no symmetry caused by precession (i.e., no point-symmetry).  

In PNe, such lack of any kind of symmetry might result from stochastic mass transfer from the giant progenitor of the PN to the stellar companion that launches the jets, or from the presence of a third star in the system (e.g. \citealt{BearSoker2017a}). In CCSNe, jittering jets lead to an ejecta with no symmetry. Rather than the effect of binary or triple stellar evolution, the jittering jets result from perturbations in the pre-collapse core, like convection, that are amplified by instabilities during the collapse and lead to intermittent accretion disks that launch the jittering jets. Despite these differences between PNe and CCSNe in the mechanisms that cause the change in the jets’ directions, the morphologies of some PNe might hint on the `messy' structure that is expected by jittering jets in some cases. 

\begin{figure}
 \centering
\hskip -2.00 cm
\includegraphics[trim= 2.4cm 0.0cm 0.0cm 0.0cm,clip=true,width=0.8 \textwidth]{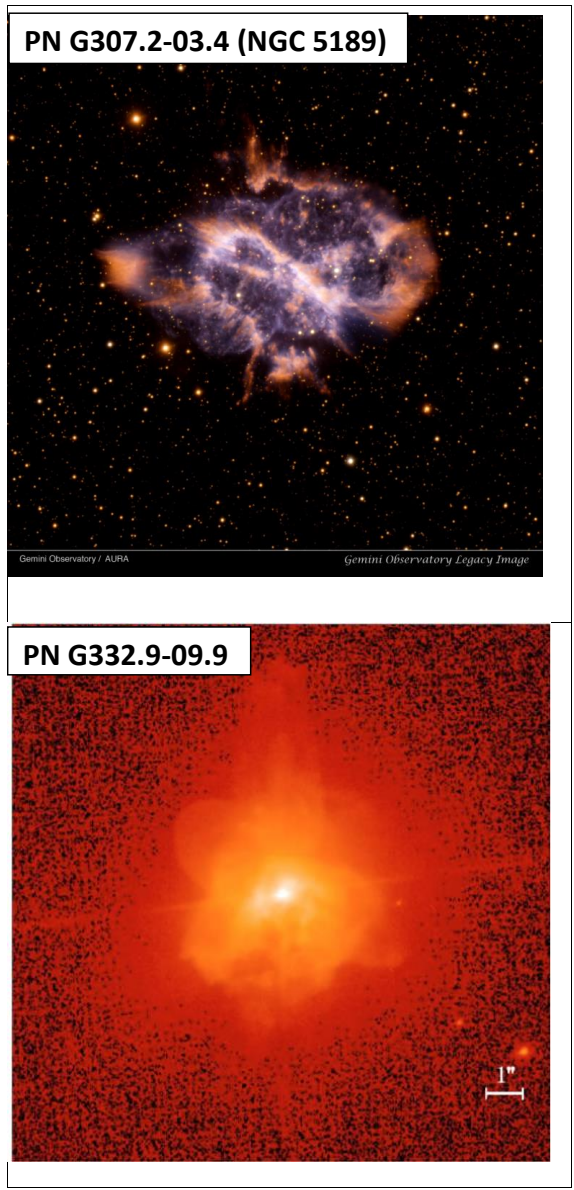} \\
 \vskip -1.50 cm
 \caption{Two PNe that are thought to be shaped by jets but lack any kind of symmetry. These can hint on the outcome of jittering jets in some CCSNe. The upper panel shows PNG~307.2-03.4 (NGC 5189) that we take from Gemini (credit: Gemini Observatory/AURA), the lower panel shows PNG~332.9-09.9 (Hen~3-1333; CPD-56◦8032) that we take from \cite{Chesneauetal2006}. }
 \label{PNe}
 \end{figure}

In Fig. \ref{PNe} we present two PNe that lack symmetry, PNG~307.2-03.4 (NGC~5189) and PNG~332.9-09.9 (Hen~3-1333; CPD-56◦8032). Several studies (e.g. \citealt{Goncalvesetal2001, Sahaietal2011, Sabinetal2012, Manicketal2015}) suggest that the major shaping of NGC~5189 is by jets which were launched in more than one episode and with varying directions. 
 The lobes in PNG~332.9-09.9 (e.g., \citealt{Sahaietal2011, DanehkarParker2015}) suggest to us (also \citealt{Sahaietal2011}) that this PN was also shaped by jets. We attribute the lack of symmetry in each of these PNe to a triple stellar system \citep{BearSoker2017a}. The major shaping is by jets, but the tertiary star affects the location of the secondary star that launches the jets and/or the direction of the jets to vary over time (e.g., \citealt{AkashiSoker2017}).

These two PNe (and there are more examples) show that shaping by jets can lead to an expanding nebula, a PN or a SN ejecta, that lack any clear symmetry. In the present study we suggest that this is the case for SNR 1987A.
  
\section{Discussion and Summary}
\label{sec:summary}
 
We examined and discussed the morphology of SNR~1987A (section \ref{sec:1987A}) in comparison to four SNRs that have clumpy ejecta and signatures of shaping by jets (section \ref{sec:Clump}). Instabilities alone cannot account for the clumpy ejecta of SNR~1987A \citep{Soker2017a, Abellanetal2017}). Although in SNR~1987A there is no direct evidence for jets, we argued in section \ref{subsec:Properties} that SNR~1987A shares some key properties with the four SNRs that we described in section \ref{sec:Clump}, and from that we strengthened the claim that jets played a crucial role in the explosion of SN~1987A.  
  
Although the comparison is qualitative, it stands on a solid ground because some of the morphological features, such as opposite ears, a barrel shape and torus like ejecta, are shared also with planetary nebulae where jets are observed. Our earlier comparison of SNRs with planetary nebulae (\citealt{BearSoker2017b, Bearetal2017, GrichenerSoker2017}) and the comparison of SNR~1987A to some SNRs, bring us to support the claim that stochastic jets shaped and powered SNR~1987A, namely, the jittering jets mechanism. In section \ref{sec:PNe} we present two PNe that suggest that jittering jets can lead to a morphology that lack any clear symmetry. 

In section \ref{subsec:fluctuations} we made a preliminary attempt to quantify the fluctuations in the angular momentum of the mass that is accreted on to the newly born NS. We started with the fluctuations in radial momentum of the ejected mass that lead to a net radial momentum of the ejecta. This net radial momentum is opposite to that of the NS that suffers a natal kick. We assumed that the accreted mass suffers similar fluctuations at $r \simeq 100 \km$ from the NS, and that the fluctuations are similar in radial and tangential directions. This implies that material cannot be accreted along the polar directions, and that two opposite cones along the polar directions with an opening angle of $\theta_a \simeq 30^\circ$ (equation \ref{eq:alphaA}) have very low accretion flux. In the jittering jets explosion mechanism jets are launched along these cones.
 
 In several SNRs including SNR~1987A the two opposite outer parts of the elongated structure are not equal. We attributed this (section \ref{subsec:asymmetryjets}) to the nature of the jittering jets explosion mechanism. The duration of each jet-launching episode lasts for about $t_{\rm jet} \simeq 0.05-0.1 \s$. The relaxation time of the accretion disk/belt is the viscous time that amounts to $t_{\rm visc} \simeq 0.01-0.1 \s$ (equation \ref{eq:tvisc1}). Since $t_{\rm visc}$ is not much shorter than $t_{\rm jet}$, the disk/belt does not reach a complete relaxation during the jets launching episode.  Hence, the two sides of the accretion disk/belt might not be equal. As well, the two opposite cones through which the jets are launched might not be equal in size and in the inward momentum flux inside them. A larger inward momentum flux inside the cone makes the outgoing jet weaker. 
 
We end this study by reiterating our earlier call (e.g. \citealt{Papishetal2015, Soker2017a, Bearetal2017}), but this time more loudly, for a paradigm shift from neutrino-driven explosion to a jet-driven explosion of CCSNe. 
  
We thank Avishai Gilkis for helpful comments. This research was supported by the Asher Fund for Space Research at the Technion, and the Israel Science Foundation.

\end{document}